\documentclass[5p]{elsarticle}

\biboptions{sort&compress}

\date{\today}

\usepackage{graphicx}
\usepackage{xcolor}
\usepackage{amssymb}
\usepackage{amsmath}
\usepackage[switch]{lineno}
\usepackage{setspace}
\usepackage{longtable}
\usepackage{caption}
\usepackage{subcaption}
\usepackage{comment}

\usepackage{hyperref}

\nolinenumbers
\journal{Chemical Engineering Science}

\begin{document}
	
	\begin{frontmatter}
		
		\title{\textcolor{black}{A network-based approach to measure granule size distribution for discrete element modeling of granulation}}
		
		\author[aff1]{Shubham Jain}
		
		\author[aff1]{\texorpdfstring{Anurag Tripathi \corref{cor1}}{Anurag Tripathi}}
		\ead{anuragt@iitk.ac.in}
		
		\author[aff2]{Jayanta Chakraborty}
		
		\author[aff3]{Jitendra Kumar}
		
		\cortext[cor1]{Corresponding author}
		
		\affiliation[aff1]{
			organization={Department of Chemical Engineering},
			addressline={Indian Institute of Technology},
			city={Kanpur 208016},
			country={India}}
		
		\affiliation[aff2]{
			organization={Department of Chemical Engineering},
			addressline={Indian Institute of Technology},
			city={Kharagpur 721302},
			country={India}}	
		
		\affiliation[aff3]{
			organization={Department of Mathematics},
			addressline={Indian Insitute of Technology},
			city={Ropar 140001},
			country={India}}

		\begin{abstract}
            Drum granulation is a size enlargement process where granular material is agitated with a liquid binder to form larger size granules. Discrete element modeling is increasingly being used to better understand and investigate the granulation process. However, unlike experiments the measurement of granule size within a DEM framework often necessitates an explicit quantitative definition of a granule and a corresponding granule identification method. In this work, we show that the existing definitions and the associated methods in literature are ineffective at identifying granules for dense flows such as during drum granulation. We propose an improved definition and granule identification method based on community-detection used in network science literature. The proposed method better identifies granules in a drum granulator as benchmarked against liquid-settling. We also vary granulation process parameters like liquid content and fill level and study their effect on the cumulative granule size distribution attained after drum granulation. We find that the existing granule-identification methods fail to reproduce the well-known effects of process parameters on the cumulative granule size distribution. The proposed method, based on community detection, reproduces the effects with better accuracy.
		\end{abstract}
		
		%
		
		

	\end{frontmatter}

\section{Introduction}
\label{sec:1}
Wet granulation is a size-enlargement process which is used to improve bulk-level properties of granular material like flowability, and particle-level properties like size, strength and density \cite{litster_science_2004}. 
In wet granulation, granular material is agitated along with a liquid binder. This process results in the particles agglomerating due to the surface-tension and viscous force of the liquid binder to form larger particles called granules \cite{shabanian_improved_2020}. This process is performed \textcolor{black}{commonly across different industries using} equipments like high-shear mixers \cite{gantt_high-shear_2005,gantt_determination_2006,hassanpour_analysis_2013,sarkar_dem_2018,tamrakar_dem_2019,you_investigation_2021,you_comparisons_2022}, fluidized-beds\cite{goldschmidt_discrete_2003,kafui_fully-3d_2008,mansourpour_investigating_2014,boyce_analysis_2017,boyce_growth_2017,zhu_cfd-dem_2023} and rotating drums \cite{mishra_preliminary_2002,bhimji_discrete_2009,liu_dynamics_2011,lin_study_2016,behjani_investigation_2017,vo_agglomeration_2019,baba_dempbm_2021,das_efficient_2022,shi_numerical_2022,nakamura_numerical_2022,dong_partially_2023,wang_numerical_2023,wang_effect_2024,adetayo_effect_1993,iveson_fundamental_1996,iveson_fundamental_1998,iveson_growth_1998,walker_drum_2000,wauters_growth_2002,wauters_liquid_2002,sastry_investigation_2003,heim_effect_2004,pan_granule_2006,degreve_spray-agglomeration_2006,ramachandran_experimental_2008,ahmadian_analysis_2011,chou_experimental_2011,nosrati_drum_2012,lefebvre_erosion_2013,xue_effect_2013,bowden-green_investigation_2016,ileleji_experimental_2016,probst_effect_2016,rodrigues_drum_2017,briens_comparison_2019,chen_production_2019,briens_comparison_2020,jop_wet_2021,fazzalari_investigation_2024,wang_multi-form_2007,glaser_model_2009,poon_experimental_2009,ramachandran_model-based_2012}. The choice of equipment depends on the desired granule attributes (size, shape, density), operation scale and the overall process circuit. Rotating drums are a popular choice of equipment which provide very high throughputs and produce high density, spherical granules, albeit suffering from large recycle ratios \cite{litster_science_2004}. \\
Material properties such as binder surface tension, binder viscosity, initial particle size distribution, contact angle and process parameters such as liquid content, drum size, rotation speed, and spraying method affect the granular flow \cite{govender_granular_2016,jarray_wet_2019,orozco_rheology_2020}, granulation rate-processes \cite{iveson_fundamental_1996,iveson_fundamental_1998,iveson_growth_1998} and granule attributes \cite{mishra_preliminary_2002,bhimji_discrete_2009,liu_dynamics_2011,lin_study_2016,behjani_investigation_2017,vo_agglomeration_2019,baba_dempbm_2021,das_efficient_2022,shi_numerical_2022,nakamura_numerical_2022,dong_partially_2023,wang_numerical_2023,wang_effect_2024,adetayo_effect_1993,iveson_fundamental_1996,iveson_fundamental_1998,iveson_growth_1998,walker_drum_2000,wauters_growth_2002,wauters_liquid_2002,sastry_investigation_2003,heim_effect_2004,pan_granule_2006,degreve_spray-agglomeration_2006,ramachandran_experimental_2008,ahmadian_analysis_2011,chou_experimental_2011,nosrati_drum_2012,lefebvre_erosion_2013,xue_effect_2013,bowden-green_investigation_2016,ileleji_experimental_2016,probst_effect_2016,rodrigues_drum_2017,briens_comparison_2019,chen_production_2019,briens_comparison_2020,jop_wet_2021,fazzalari_investigation_2024,wang_multi-form_2007,glaser_model_2009,poon_experimental_2009,ramachandran_model-based_2012} in a complex way.
Thus, the design of granulation process \cite{glaser_model_2009,ramachandran_model-based_2012,walls_towards_2022} requires understanding the effect of these input parameters on the granular flow, the granulation rate processes and ultimately on the granule attributes.
While experimental investigations \cite{adetayo_effect_1993,iveson_fundamental_1996,iveson_fundamental_1998,iveson_growth_1998,walker_drum_2000,wauters_growth_2002,wauters_liquid_2002,sastry_investigation_2003,heim_effect_2004,pan_granule_2006,degreve_spray-agglomeration_2006,ramachandran_experimental_2008,ahmadian_analysis_2011,chou_experimental_2011,nosrati_drum_2012,lefebvre_erosion_2013,xue_effect_2013,bowden-green_investigation_2016,ileleji_experimental_2016,probst_effect_2016,rodrigues_drum_2017,briens_comparison_2019,chen_production_2019,briens_comparison_2020,jop_wet_2021,fazzalari_investigation_2024} have provided valuable insight into these effects, measurement difficulties have limited multi-scale understanding. In particular the effect of particle-particle interactions on the granulation rate processes and granule attributes is difficult to investigate via experimental means. In recent years, progress in discrete element modeling (DEM) \cite{cundall_discrete_1979} of granulation  has allowed for such investigations. While DEM provides the most detailed description of the process, the identification of granules and measurement of granule size distribution within the DEM framework is not straightforward. 

\begin{figure}[h]
	\centering
	\includegraphics[width=8cm]{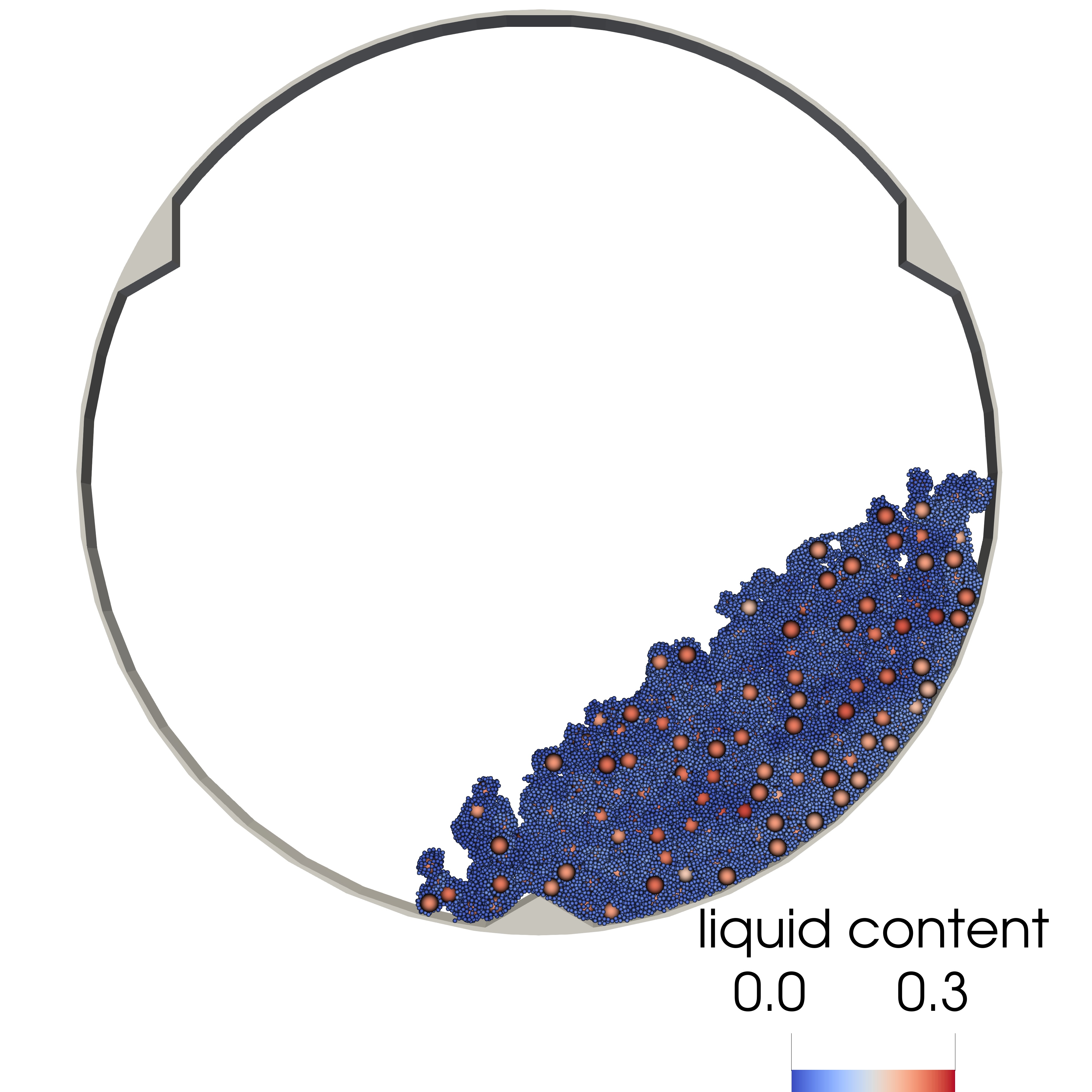}
	\caption[\relax]{Snapshot showing granulation in a drum \textcolor{black}{for} fill level \texorpdfstring{$f=16\%$}{f=16\%} and the \textcolor{black}{total} liquid content \texorpdfstring{$L=12\%$}{L=12\%}.}
	\label{fig:schematic_granulation}
\end{figure}

The earliest attempts at modeling granulation using DEM were made by Tardos et al. \cite{khan_stability_1997,talu_computer_2000}. They performed 2D DEM simulations of a mass of dry and wet particles in a constant shear field. The wet particles interacted with each other and with dry particles via capillary and viscous forces due to their binder layers. To identify granules, they used an image-processing approach. They plotted the particles in their system by coloring them black for wet particles and grey for dry particles on a white background. \textcolor{black}{By removing all the dry particles from the snapshots except those completely enclosed by wet particles, and using a pattern recognition technique, the authors identified the granules and measured their size and shape.} Thus, a group of wet particles and enclosed dry particles bound by viscous and surface tension forces was considered as a granule in their work. 

Gantt and Gatzke \cite{gantt_high-shear_2005} modeled a high-shear granulator using DEM. They used physics-based coalescence criteria developed by Litster et al. \cite{litster_science_2004} to model the agglomeration process. \textcolor{black}{Two} colliding particles were considered to have agglomerated if their viscous Stokes number (ratio of inertia and viscous dissipation) was below a critical value. Such colliding particles were then replaced by a single spherical particle on the basis of mass and volume conservation. Thus, they implicitly defined a granule as a larger particle formed by \emph{successful} (i.e., completely dissipative) collisions of two or more particles. This framework does not require a granule identification method like Tardos et al. \cite{khan_stability_1997,talu_computer_2000}. However, the replacement of agglomerating particles by a single spherical particle leads to loss of important morphological attributes of the granules. Goldschmidt et al. \cite{goldschmidt_hydrodynamic_2001,goldschmidt_discrete_2003} studied agglomeration in fluidized bed. In their work, two wet particles or a wet and a dry particle rebounded or coalesced with each other depending on a probability of their wet surface area colliding with each other. The coalesced particles were replaced by a single bigger particle on the basis of volume, mass, and momentum conservation. Similar to Gantt and Gatzke \cite{gantt_high-shear_2005}, granules could be readily identified in their work. \textcolor{black}{However,} these approaches based on coalescence criteria of two particle collision are not readily applicable for dense systems like drum granulators where the average coordination number is much higher than 2.

In contrast to the above two approaches, Thornton et al. \cite{mishra_preliminary_2002,kafui_fully-3d_2008,bhimji_discrete_2009} performed wet granulation simulations using DEM in rotating drum as well as fluidized bed by using the concept of surface energy as described in the JKR theory \cite{johnson_surface_1971}. While they \cite{mishra_preliminary_2002,kafui_fully-3d_2008,bhimji_discrete_2009} reported granule size distribution in their work, they did not explicitly mention how they identified an agglomerate and its boundary particularly in a dense system like the rotating drum. Hassanpour et al. \cite{hassanpour_analysis_2013,behjani_investigation_2017} also used JKR theory to model seeded granulation in rotating drum as well as high-shear granulator. For the purpose of quantitative analysis, they defined a seeded granule as a seed particle bonded via cohesive force with fines irrespective of the number of fines \textcolor{black}{\cite{hassanpour_analysis_2013}}. Thus, a single seed bonded with a single fine was also counted as a seeded granule. According to this definition the number of seeded granules reduced with increasing rotation speed. Interestingly, this trend was reversed when they counted only the seeds with at least 50\% of their surface covered as seeded granules. This clearly exemplifies the importance of granule definition in DEM in order to get accurate insight into the granulation process.

Tamrakar et al. \cite{tamrakar_dem_2019} investigated granulation in a high-shear mixer using DEM. They did not identify the granules and their size from their DEM simulations \textcolor{black}{and} used the average number of liquid bridges as a measure of the granulation performance. You et al. \cite{you_investigation_2021,you_comparisons_2022} performed experiments and simulations of the iron-ore fine granulation process in horizontal high-shear granulator. They used the liquid bridge \textcolor{black}{force} to model the effect of binder and thus the rate processes like coalescence and breakage in their work. \textcolor{black}{Similar to previous investigations \cite{hassanpour_analysis_2013,behjani_investigation_2017, tamrakar_dem_2019}, they used proxy measures like average number of fines per seed, and average liquid-bridge force to quantify the granulation performance and did not report granule size distribution.} Sarkar and Chaudhuri \cite{sarkar_dem_2018} also studied granulation in high-shear granulator using liquid bridge model in DEM \textcolor{black}{and} used the number of liquid bridges and the coordination number of every particle at any instant as a measure of the granulation process.
 
Shi et al. \cite{shi_numerical_2022} used the liquid bridge model to study the iron-ore granulation process in a rotating drum. They reported the cumulative granule size distribution in their work by performing sieving of the particles in DEM. For this they used the sieve function in LIGGGHTS which passes particles through the sieve by calculating a probability as a ratio of the particle area and the sieve spacing area \cite{noauthor_fix_nodate}. It is unclear how accurately this function models the actual sieving process. \textcolor{black}{Using a different sieving process from Shi et al.'s  \cite{shi_numerical_2022} probability-based sieving approach, Wang et al. \cite{wang_numerical_2023,wang_effect_2024} separated out granules and computed their surface coverage but did not report the cumulative granule size distribution}. 
 
Boyce et al. \cite{boyce_growth_2017} simulated the growth and breakage of a wet agglomerate in a fluidized bed \textcolor{black}{and} considered a group of particles interconnected with liquid-bridges as a granule. Vo et al. \cite{vo_agglomeration_2019} used the liquid bridge model to investigate the accretion and erosion of particles from a single granule as it moves in a rotating drum with wet and dry particles. Importantly, in their model only wet particles could form liquid bridges with each other. Furthermore, they were only interested in tracking the size of the single initial granule. Thus they simply tracked the wet particles that form a liquid bridge with the single initial granule to determine accretion \textcolor{black}{and} erosion events. Thus, a group of wet particles interconnected with liquid-bridges was considered as a granule. Nakamura et al. \cite{nakamura_numerical_2022} used liquid bridge force to model granulation in a rotating drum using \textcolor{black}{discrete element method and population balance modeling (DEM-PBM)}. Within a DEM-PBM framework, they considered particle-pairs as having agglomerated if they had a liquid-bridge present between them and if the relative motion between the two particles was below a threshold value. Thus, implicitly they defined a granule as two or more particles connected with each other through binary pendular liquid bridges and having nearly zero relative motion. Zhu et al. \cite{zhu_cfd-dem_2023} performed CFD-DEM simulations of fluidized bed. To detect the agglomerates formed in this system, they used an agglomerate detection algorithm based on the contacts between different pairs of overlapping particles and the common particles found in these pairs. They considered particle-pairs with common particles as part of a single agglomerate. They repeated this process for all the common particles found in the particle-pairs. Thus, in their work, they defined a granule as a group of overlapping particles in contact with each other. \textcolor{black}{The methods used by various authors are briefly reviewed below and are summarized in Table \ref{table:1}.}

\begin{table*}[h!]
	\centering
	\caption{Literature on discrete element modeling of granulation.}
	\label{table:1}
	
	\begin{tabular}{p{2 cm}p{2 cm}p{4 cm}p{4 cm}p{4 cm}}
		\hline
		Author & System & Wetting-nucleation method & Agglomeration-breakage method & Granule identification method\\ \hline
		Talu et al. \cite{talu_computer_2000} & Shear flow & Randomly distributed wet particles & Liquid-bridge force & Wet particles in contact with each other identified via image-processing\\ \hline
		Muguruma et al. \cite{muguruma_numerical_2000} & Centrifugal tumbling granulator & Not mentioned & Liquid-bridge force (neglected lubrication force) & Not measured. They only investigated kinematics. \\ \hline
		Mishra et al. \cite{mishra_preliminary_2002} & Drum & Wetting zone & JKR \cite{johnson_surface_1971} & Not mentioned explicitly \\ \hline
		Goldschmidt et al. \cite{goldschmidt_discrete_2003} & Fluidized bed & Discrete particles as droplets & Replacement of colliding particle-pairs with a single particle based on a collision criteria & Identification not required \\ \hline
		Gantt and Gatzke \cite{gantt_high-shear_2005} & High-shear granulator & Same liquid content to all particles & Replacement of colliding particle-pairs with a single particle based on a collision criteria & Identification not required \\ \hline
		Kafui and Thornton \cite{kafui_fully-3d_2008} & Fluidized bed & Wetting zone & JKR & Not mentioned explicitly \\ \hline
		Bhimji \cite{bhimji_discrete_2009} & Drum & Wetting zone & JKR & Not mentioned explicitly \\ \hline
		Hassanpour et al. \cite{hassanpour_analysis_2013} & High-shear granulator & Same surface energy to all particles & JKR & A large seed covered by fines with a minimum surface-coverage fraction. \\ \hline
		Mansourpour et al. \cite{mansourpour_investigating_2014} & Fluidized bed & Not wet granulation & Solid-bridge force and replacement by multisphere & Not needed \\ \hline
		Behjani et al. \cite{behjani_investigation_2017} & Drum (Continuous) & Same surface energy to all particles & JKR & Not measured \\ \hline
		Boyce et al. \cite{boyce_growth_2017,boyce_analysis_2017} & Fluidized bed & Single wet granule surrounded by a bed of dry particles & Liquid-bridge force & Common contact with liquid bridge presence \\ \hline
		Sarkar and Chaudhuri \cite{sarkar_dem_2018} & High-shear granulator & All particles were assigned the same liquid content & Liquid-bridge force & Not measured \\ \hline
		Vo et al. \cite{vo_agglomeration_2019} & Drum & Randomly distributed wet particles and an initial granule of wet particles & Liquid-bridge force & Wet particles in contact via liquid-bridge. Contact between wet and dry particles were neglected. \\ \hline
		Tamrakar et al. \cite{tamrakar_dem_2019} & High-shear granulator & Discrete particles as droplets & Liquid-bridge force & Not measured \\ \hline
		You et al. \cite{you_investigation_2021} & High-shear granulator & Same liquid content to all particles & Liquid-bridge force & Not measured \\ \hline
		Shi et al. \cite{shi_numerical_2022} & Drum & Spray zone & Liquid-bridge force & Sieving \\ \hline
		Nakamura et al. \cite{nakamura_numerical_2022} & Drum & Uniform-binder & Liquid-bridge force & Particles in contact via liquid-bridge having nearly zero relative motion between them\\ \hline
		Zhu et al. \cite{zhu_cfd-dem_2023} & Fluidized-bed & Discrete particles as droplets & Liquid-bridge force & Particles in contact via liquid-bridge \\ \hline
		Wang et al. \cite{wang_numerical_2023} & Drum & Wetting zone & Liquid-bridge force & Sieving \\
		\hline
	\end{tabular}
\end{table*}

In summary, DEM investigations of granulation in general and drum granulation in particular have not focused much on the process of granule identification and measurement of granule size distribution within the DEM framework. Many studies on granulation \cite{muguruma_numerical_2000, behjani_investigation_2017, sarkar_dem_2018, tamrakar_dem_2019, you_investigation_2021} did not measure the granule size and granule size distribution. They used proxy measures like average coordination number and average liquid bridge force as a measure of granulation performance. A few studies \cite{mishra_preliminary_2002, kafui_fully-3d_2008, bhimji_discrete_2009} measured granule size but did not explicitly report the measurement method. Most studies which measured the granule size and/or \textcolor{black}{their distribution} used a simple definition of a granule and corresponding method to identify granules \cite{talu_computer_2000,boyce_analysis_2017,boyce_growth_2017,vo_agglomeration_2019,nakamura_numerical_2022,zhu_cfd-dem_2023}. This method, named here as the component detection method, is discussed in detail in the section \ref{sec:4}. In this work, we \textcolor{black}{investigate} the adequacy of the component-detection method in granule identification and measurement of granule size distribution by comparing it against established methods like sieving and liquid-settling. To this end, we \textcolor{black}{perform} granulation in a rotating drum and visualized the granules identified by the component detection method, sieving and liquid-settling. We observed that the granules identified by the component-detection method were either too large encompassing nearly all the wet granular mass in the drum or were too small (singlets and doublets) as compared to the benchmark methods. We also \textcolor{black}{vary} two \textcolor{black}{important} granulation process parameters (liquid content and fill level) and \textcolor{black}{study} their effect on cumulative granule size distribution as measured by the component detection method, sieving and settling. We \textcolor{black}{find} that the component detection method \textcolor{black}{is} not able to capture the well-known effects \textcolor{black}{of} the process parameters on the cumulative granule size distribution produced after drum granulation. Therefore, we \textcolor{black}{propose} a new method of granule identification based on the community detection technique from the network science literature \cite{newman_networks_2018}. Visualization of the granules identified by the community detection method and \textcolor{black}{their} comparison against established methods \textcolor{black}{demonstrate} its effectiveness in identifying granules \textcolor{black}{from DEM simulations} of drum granulation. We \textcolor{black}{are} also able to reproduce the well-known effects of the process parameters on the drum granulation process by measuring the cumulative granule size distribution using the proposed method.  
		
\section{\textcolor{black}{Simulation methodology}}
\label{sec:2}

In this work, the motion of particles in a drum during granulation \textcolor{black}{is} modeled using \textcolor{black}{discrete element method (DEM)} \cite{cundall_discrete_1979}. DEM is a Lagrangian approach in which the positions of a collection of discrete particles are evolved in time using Newton’s laws of motion \cite{norouzi_coupled_2016},

\begin{equation}
	\label{eq:1}
	m_i\frac{\mathrm{d}^2\vec{x}_i}{\mathrm{d}t^2} = \vec{F}_i, 
\end{equation} 

\begin{equation}
	\label{eq:2}
	I_i\frac{\mathrm{d}\vec{\textcolor{black}{\omega}}_i}{\mathrm{d}t} = \vec{M}_i.
\end{equation} 

Here $\vec{x}_i$ is the position and \textcolor{black}{$\vec{\omega}_i$ is the angular velocity} of the particle. The subscript $i$ refers to the particle ID. $\vec{F}_i$ is the net force and $\vec{M}_i$ is the net torque acting on particle $i$, while $m_i$ and $I_i$ are the mass and moment of inertia of particle $i$ respectively.

For wet granulation in a rotating drum, the net acceleration of the particles is due to the gravitational force as well as particle-particle, particle-wall, and fluid-particle interactions \cite{litster_science_2004}. These interactions are taken into account in DEM as

\begin{equation}
	\label{eq:3}
	\vec{F}_i = m_i\vec{g} + \sum_{j,\, j\neq i} \vec{F}_{ij} 
\end{equation}

\begin{equation}
	\label{eq:4}
	\vec{M}_i = \sum_{j,\, j\neq i} \vec{M}_{ij} + \vec{M}_{f,i} 
\end{equation}

where $\vec{g}$ is the acceleration due to gravity, \textcolor{black}{$\vec{F}_{ij} = \vec{F}_{c,ij} + \vec{F}_{lb,ij}$ is the net force between particle $i$ and particle $j$, \textcolor{black}{where} $\vec{F}_{lb,i}$ is the liquid bridge force and $\vec{F}_{c,ij}$ is the particle-particle contact force between particles $i$ and $j$}. Similarly, $\vec{M}_{ij}$ is the net torque due to interactions of particles $i$ and $j$, and $\vec{M}_{f,i}$ is the net torque due to the fluid-particle interactions. 
 
\textcolor{black}{The particles are modeled as deformable spheres that overlap when they come in contact and stay in contact for a small finite time. The} particle-particle contact force is resolved using force-displacement models which depend on the normal and tangential overlaps between particles. A non-linear viscoelastic contact model namely the Hertz-Mindlin model \cite{norouzi_coupled_2016} is used in the present work to model the particle-particle contact force. The particle-wall interactions are modeled using the particle-particle approach by discretising the walls into small elements and considering each wall element as a particle with infinite radius. Thus the contact force is expressed as

\begin{equation}
	\label{eq:7}
	\vec{F}_{c,ij} = \vec{F}_{cn,ij} + \vec{F}_{ct,ij}
\end{equation}

where the subscripts $n$ and $t$ refer to the normal and tangential components of the contact force. The normal and tangential \textcolor{black}{force} components are \cite{norouzi_coupled_2016}

\begin{multline}
	\label{eq:8}
	\vec{F}_{cn,ij} = \frac{4}{3} \, Y^*\sqrt{r^*\delta_n} \,  \, \delta_n \, \vec{n}_{ij} \\ + 2 \, \sqrt{\frac{5}{6}} \, \frac{\ln{e}}{\sqrt{\ln{e}^2 + \pi^2}} \, \sqrt{2Y^*m^*\sqrt{r^*\delta_n}} \, \vec{v}_{n,ij},
\end{multline}

\begin{multline}
	\label{eq:9}
	\vec{F}_{ct,ij} = 8 \, G^*\sqrt{r^*\delta_n} \, \delta_t \vec{t}_{ij}  \\ + 2 \, \sqrt{\frac{5}{6}} \, \frac{\ln{e}}{\sqrt{\ln{e}^2 + \pi^2}}\sqrt{8G^*m^*\sqrt{r^*\delta_n}} \, \vec{v}_{t,ij}.
\end{multline}

Here $Y^*$, $r^*$, $\delta_n$, $e$, $m^*$, $G^*$, $\delta_t$, $\vec{v}_{n,ij}$ and $\vec{v}_{t,ij}$  are the effective Young's modulus, effective radius, normal overlap, coefficient of restitution, effective mass, effective shear modulus, tangential overlap, \textcolor{black}{normal} and tangential relative velocity respectively.

 The liquid bridge force is the summation of the viscous force and the capillary force due to the presence of liquid bridge between particles. In the present work, the viscous force \textcolor{black}{is} modeled according to Nase et al.\cite{nase_discrete_2001} and the capillary force according to Rabinovich et al.\cite{rabinovich_capillary_2005}. Thus, the liquid bridge force is expressed as:

\begin{equation}
	\label{eq:10}
	\vec{F}_{lb,ij} = \vec{F}_{cap,ij} + \vec{F}_{vis,ij}
\end{equation}

\textcolor{black}{The viscous force $\vec{F}_{vis,ij}$ between particles $i$ and $j$ is given as}

\begin{multline}
	\label{eq:11}
	\vec{F}_{vis,ij} = -6 \pi \mu r^* \vec{v}_{n,ij} \frac{r^*}{H}\\ -\left( \frac{8}{15}\ln\frac{r^*}{H} + 0.9588 \right) 6 \pi \mu r^* \vec{v}_{t,ij}
\end{multline}

where $\mu$ and $H$ are the binder viscosity and the separation distance between the particles respectively. \textcolor{black}{The capillary force between the grains is given as}

\begin{equation}
	\label{eq:12}
	\vec{F}_{cap,ij} = 2\pi r_{eff} \gamma_{lv} \left[ \frac{\cos(\theta)}{1 + H/2d_{ss}} + \sin(\alpha)sin(\theta+\alpha)\right], 
\end{equation}

\textcolor{black}{with $d_{ss} = \frac{H}{2} \left[ -1 +  \sqrt{1+2V/\pi r_{eff} H^2}\right]$ \textcolor{black}{where} $r_{eff}=2r_1r_2/(r_1 + r_2)$ is the effective radius, $\gamma_{lv}$, $\theta$ and $\alpha$ are the liquid-vapor surface tension, contact angle and half-filling angle, respectively.}

We study granulation in a thin slice (thickness $T = 5$ mm \textcolor{black}{in z-direction}) of \textcolor{black}{a rotating} drum with periodic boundary conditions in the z direction. The drum rotation speed is $N=36$ rpm corresponding to a Froude number $Fr=D\omega^2/2g\approx0.07$ where $\omega$ is the angular speed of the drum. A bidisperse mixture of particles \textcolor{black}{is} introduced in the drum with $d_{f}=0.5$ mm and $d_{s}=2$ mm to model the fines and seeds respectively, \textcolor{black}{so that $D/d_{min} = 200$ and $D/d_{max} = 50$}. To perform granulation, we instantaneously set \textcolor{black}{a uniform liquid content for all the seed} particles and start the rotation of the drum. The bidisperse mixture with wet particles is rotated in the drum for 10 revolutions. The drum dimensions, material properties, operating conditions and simulation parameters used in the present work are given in Table \ref{table:3}. \textcolor{black}{The wet granular mixture is then used to measure the granule size distributions using different techniques discussed below.}

\begin{table}[h]
	\centering
	\caption{Drum dimensions, material properties, operating conditions and simulation parameters used in the present work.}
	\label{table:3}
	\begin{tabular}{p{3.9 cm}p{1 cm}p{1.85 cm}}
		\hline
		Parameter & Symbol & Value\\ 
		\hline
		Drum diameter & $D$ & 100 mm \\
		Drum thickness & $T$ & 5 mm \\
		Rotation speed & $N$ & 36 rpm\\
		Fill fraction & $f$ & $\left[4\%-16\%\right]$\\
		\hline
		Particle size & $d$ & 0.5, 2 mm\\
		Particle density & $\rho$ & 2900 kg/m\textsuperscript{3} \\
		Young's modulus & $Y$ & $5.7\times10^6$ Pa\\
		Poisson's ratio & $\nu$ & 0.3 \\
		Static friction coefficient & $\mu_s$ & 0.5 \\
		Rolling friction coefficient & $\mu_r$ & 0.02\\
		Restitution coefficient & $e$ & 0.2\\
		\hline
		Surface tension & $\gamma_{lv}$ & 0.073 N/m\\
		Liquid viscosity & $\eta$ & 1 mPa s \\
		Contact angle & $\theta$ & 0\textsuperscript{o} \\
		Liquid content & $L$ & $\left[ 12\%-30\%\right]$\\
		\hline
		Timestep size & $\Delta t$ & $5\times10^{-6}$ s \\
		\hline
	\end{tabular}

\end{table}

\section{Measurement of granule size distribution}
\label{sec:4}
 
 \begin{figure*}[h]
	\centering
	\includegraphics[width=16cm]{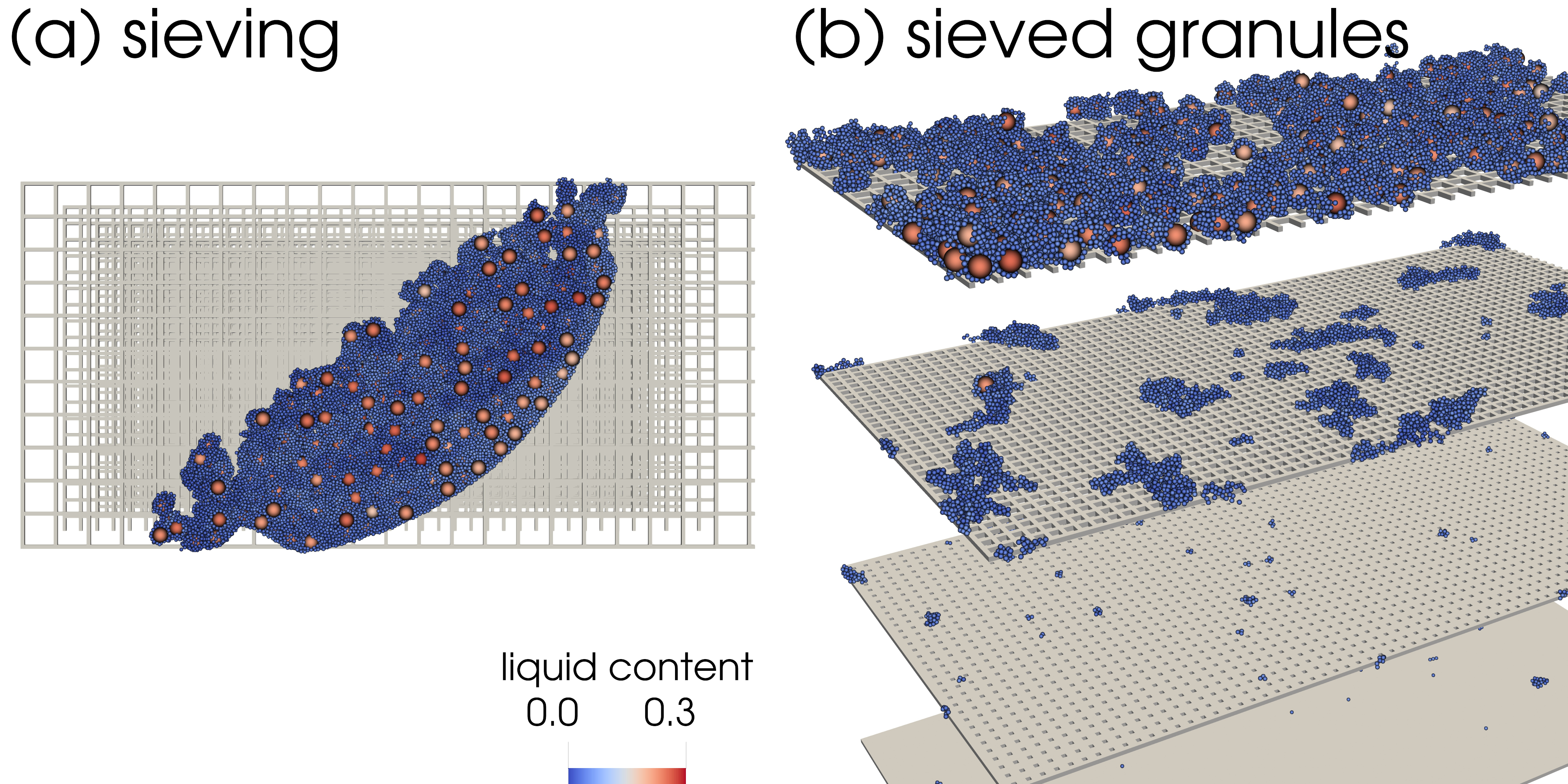}
	\caption{Snapshots of (a) \textcolor{black}{initial state before and (b) final state after the sieving for drum} fill fraction $f=0.16$ and liquid content $L=0.12$ during granulation.}
	\label{fig:schematicSieving}
\end{figure*}

\subsection{Sieving}
\label{subsec:4_2}
We \textcolor{black}{first} performed sieving \textcolor{black}{using} DEM in this work. \textcolor{black}{We used sieves of aperture size $s$ ranging from $8$ mm to $1$ mm (successive size reducing by a factor of $\sqrt{2}$)} along with a sieve of $0.5$ mm and a collection pan ($s=0$ mm) at the bottom. The \textcolor{black}{vertical} distance between each sieve was 25.4 mm according to the half-height 200 mm standard test sieves \cite{astm11standard}. The wet granular mass from the drum after granulation was gently placed on the topmost sieve $s=8$ mm \textcolor{black}{(shown in Fig.\ \ref{fig:schematicSieving}(a)) without altering the relative position of grains. All the sieves were then} wiggled in the $x, y$ and $z$ directions with sieving Froude number $Fr_s=A_s\omega_s^2/g=1.5$. Periodic boundary conditions were used in the x- and y-direction \textcolor{black}{(See Fig. \ref{fig:schematicSieving}(b))} \textcolor{black}{during the sieving process. Note that sieve meshes akin to physical sieves are used instead of the probabilistic size sieve option available in LIGGGHTS to realistically simulate the physical sieving process. The separated granules at different sieves are shown in Fig.\ \ref{fig:schematicSieving}(b).}

\begin{figure*}[h]
	\centering
	\includegraphics[width=16cm]{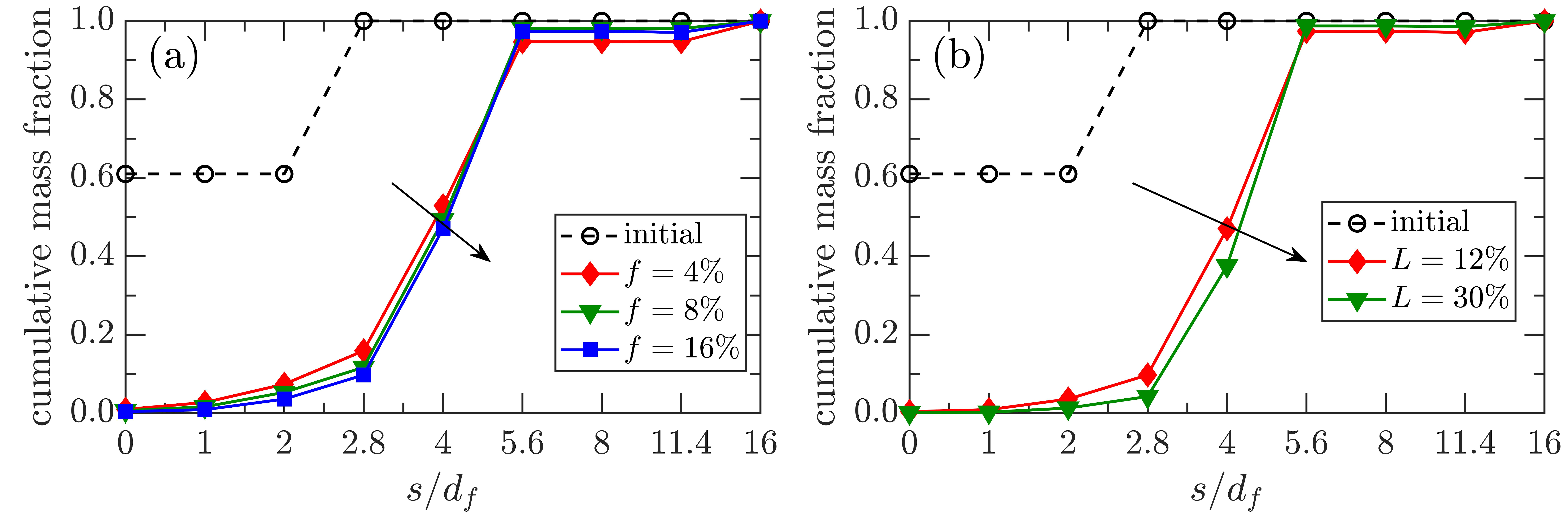}
	\caption{Cumulative granule size distribution measured by sieving showing the effect of (a) \textcolor{black}{drum fill level (for liquid content $L=12\%$) and (b) liquid content} for fill level $f=16\%$.}
	\label{fig:sieving}
\end{figure*}

Fig. \ref{fig:sieving}(a) shows the effect of fill level on the cumulative granule size distribution as measured by sieving. \textcolor{black}{The cumulative granule size distribution shifts slightly rightwards with increasing fill level, suggesting that a slightly higher proportion of larger granules are found at higher fill levels.} \textcolor{black}{A similar but much stronger effect of fill level on the cumulative granule size distribution has been reported in Ref \cite{you_comparisons_2022}.} Fig. \ref{fig:sieving}(b) shows the effect of liquid content on the cumulative granule size distribution as measured by sieving. As expected, we find a rightward shift of the cumulative granule size distribution with increasing liquid content from 12\% to 30\%. Thus, a higher proportion of larger granules \textcolor{black}{are} found when the liquid content \textcolor{black}{is} increased. \textcolor{black}{A similar trend of the cumulative granule size distribution with the increasing liquid content has been reported by \cite{chen_production_2019} in their experimental study}. Thus, we \textcolor{black}{are able to see the effect of liquid-content on the granulation process by sieving in DEM}. \textcolor{black}{The effect of fill level, however, is not observed to be very significant according to the sieving method.}

\textcolor{black}{It is important to note that current DEM models of granulation do not account for the aging of the liquid-bridge bonds and hence do not lead to formation of} stronger solid-bridges. \textcolor{black}{Therefore, performing sieving in DEM leads to significantly more breakage than experiments. Due to this reason, the expected effects of fill level \textcolor{black}{and liquid level may not be well} captured by sieving in DEM simulations. To avoid this exaggerated breakage due to vibrating sieve, we used another gentler method of liquid-settling to measure cumulative granule size distribution}.

\subsection{Liquid-settling} 
\label{subsec:4_3}
\begin{figure*}[h]
	\centering
	\includegraphics[width=16cm]{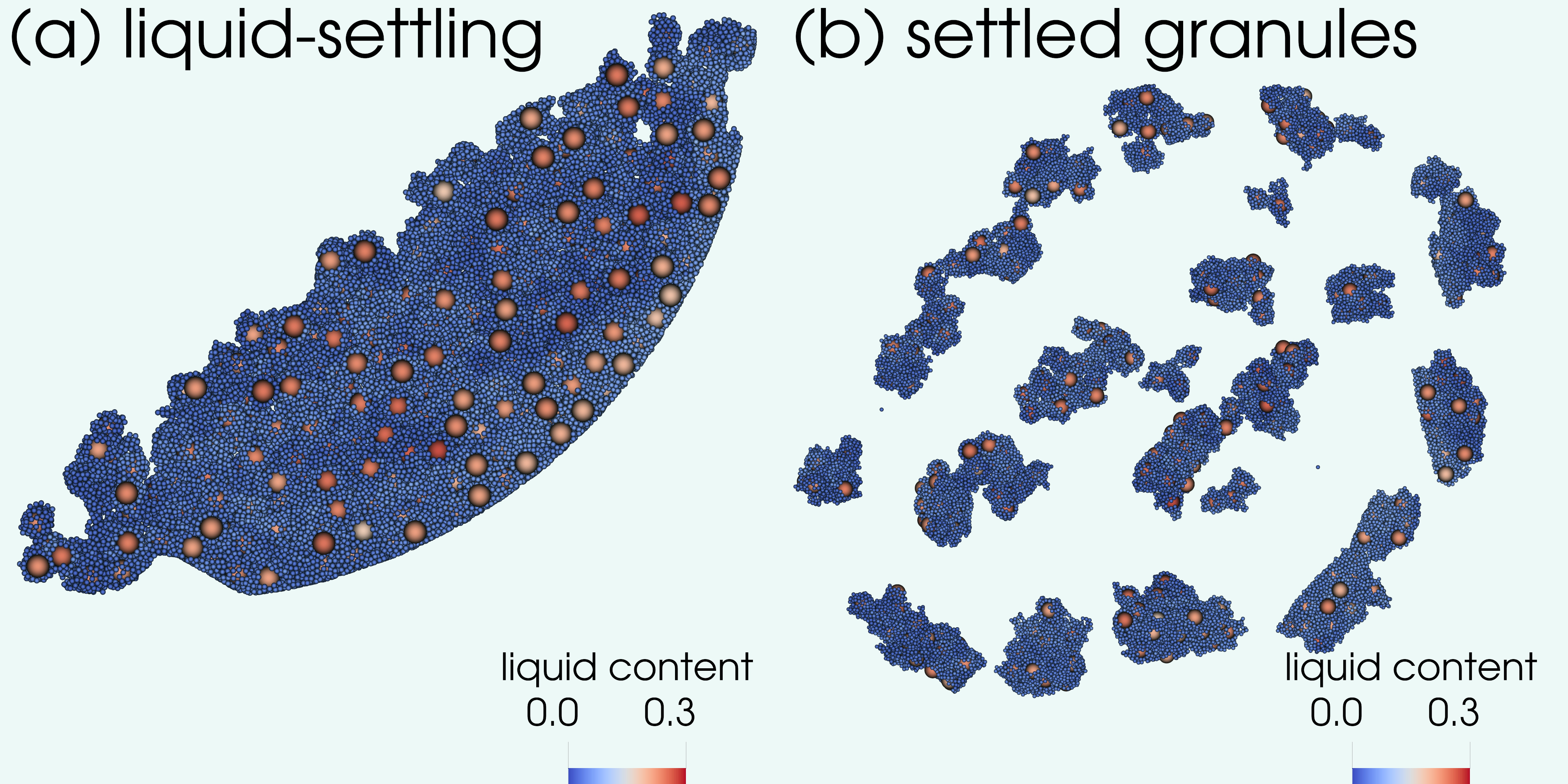}
	\caption{Snapshots of (a) \textcolor{black}{initial condition for} liquid-settling and (b) \textcolor{black}{finally separated} granules for fill fraction $f=0.16$ and liquid content $L=0.12$ during granulation.}
	\label{fig:schematicSettling}
\end{figure*}

 This size-segregation process is based on the idea that different sized granules experience different size-dependent drag force from the surrounding liquid. To simulate the effect of the drag force, we used a coupled computational fluid dynamics-discrete element method (CFD-DEM) approach for this process \textcolor{black}{using CFDEM software that couples LIGGG\-HTS and Openfoam.} The wet granular mass from the drum after granulation \textcolor{black}{is gently placed at the top of a tall liquid column containing water (see Fig. \ref{fig:schematicSettling}(a))}. The wet granular material is allowed to settle under gravity in the liquid column which leads to separation of the \textcolor{black}{entire granular mass into distinct granules. Different size granules are distributed at different locations in the column and all of these granules are shown together in Fig. \ref{fig:schematicSettling}(b))}.

\begin{figure*}[h]
	\centering
	\includegraphics[width=16cm]{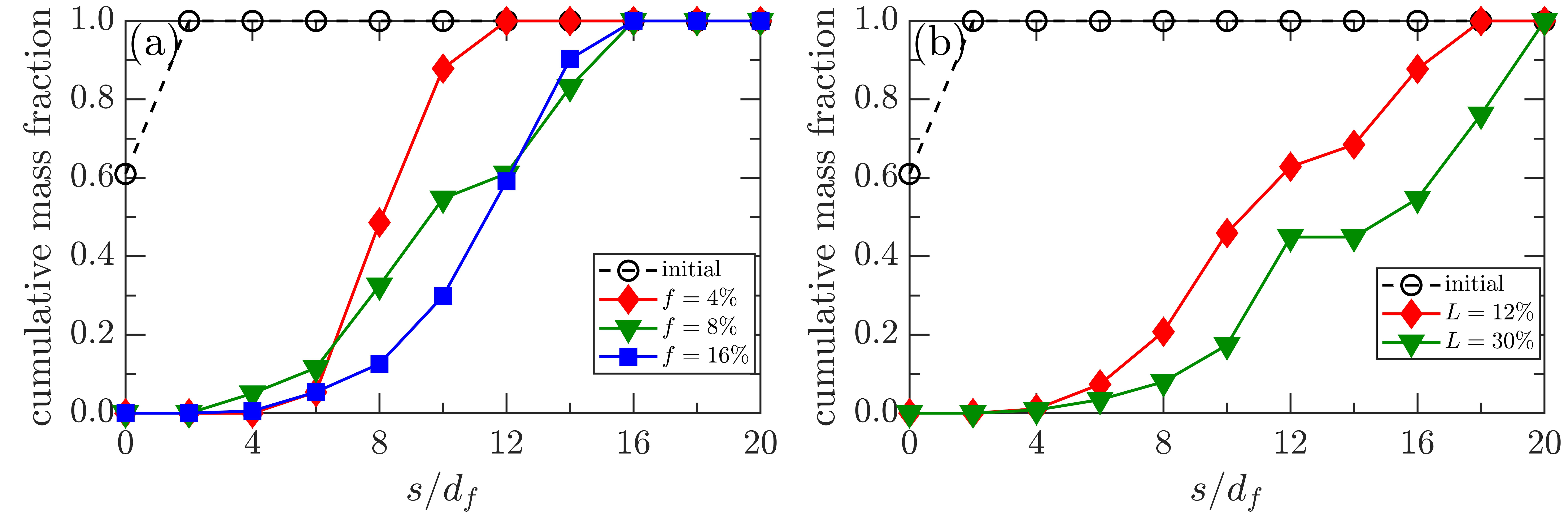}
	\caption{Cumulative granule size distribution measured by liquid-settling showing the effect of (a) fill level \textcolor{black}{for liquid content $L=12\%$} and (b) liquid content for the fill level $f=16\%$.}
	\label{fig:settling}
\end{figure*}

Fig. \ref{fig:settling}(a) shows the effect of fill level on cumulative granule size distribution as measured by liquid-settling. \textcolor{black}{Similarly, Fig. \ref{fig:settling}(b) shows the effect of liquid content. We observe a rightward shift in the cumulative granule size distribution with increasing fill level as well as with liquid content in both the cases, albeit to a larger extent compared to sieving.} 

\textcolor{black}{The effects of fill-level and liquid content through the liquid-settling method are stronger because this process is gentler as compared to the vigorous sieving process}. The excessive breakage in sieving (due to a lack of modeling bond-aging) leads to the effect of \textcolor{black}{fill level and liquid content on granulation being less pronounced as compared to liquid-settling. In the next section we} \textcolor{black}{report the results obtained from the network based methods.}

\subsection{Network-based methods}
\label{subsec:4_1}

  \begin{figure}[h]
	\centering
	\includegraphics[width=4cm]{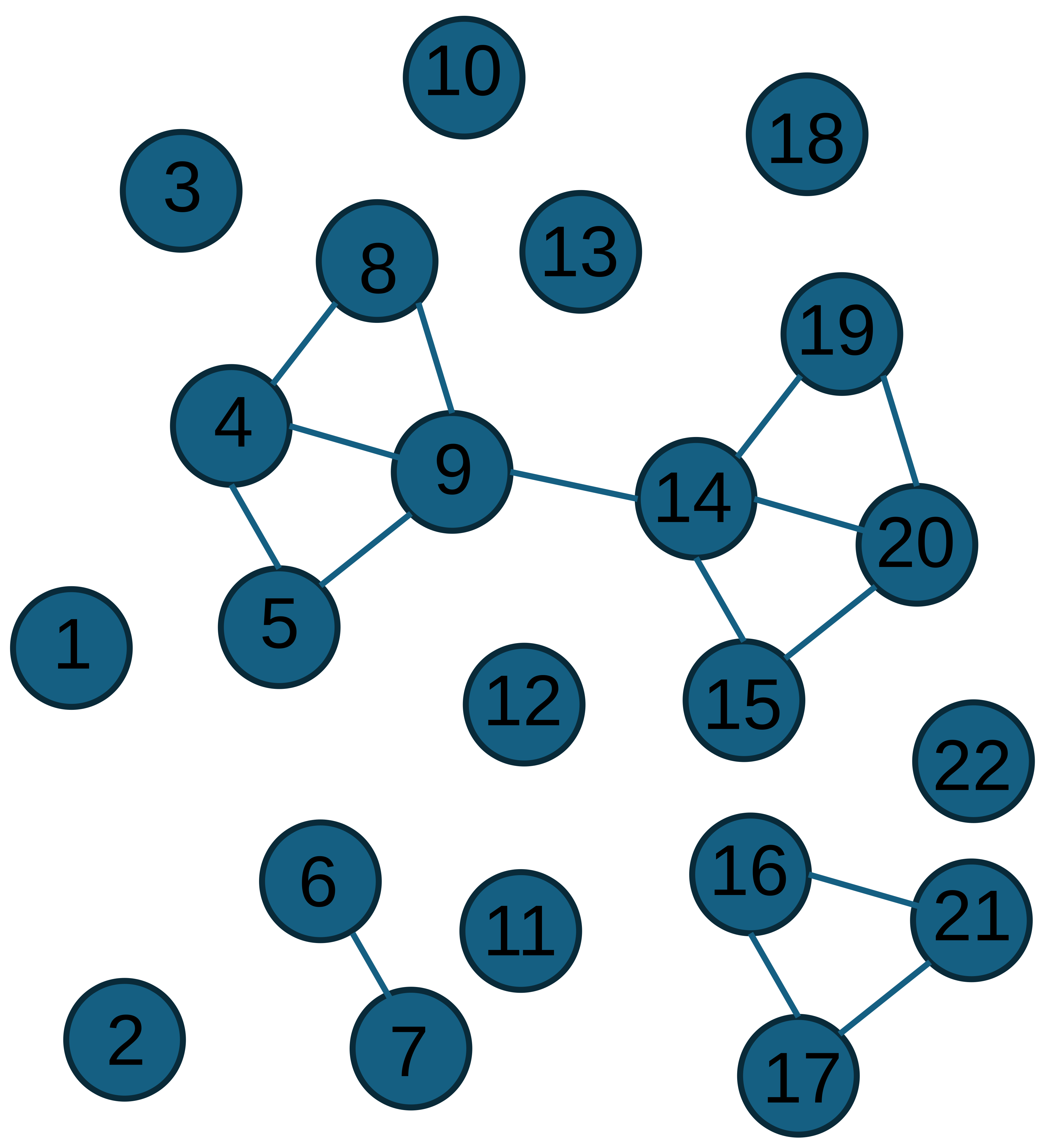}
	\caption{Schematic showing a graph with 22 nodes and 15 edges.}
	\label{fig:network}
\end{figure}
 
 The measurement of granule size distribution from discrete element modeling of drum granulation typically requires the identification of groups of particles as granules. This identification is done on the basis of quantitative definition of a granule. \textcolor{black}{To enable this we briefly introduce some basic concepts from network science \cite{newman_networks_2018} and use these to explore the granule size distribution in our system.}
 
 \subsubsection{Component-detection method}
 \label{subsubsec:4_3_1}
 
 \textcolor{black}{Let us consider a collection of particles as shown in Fig. \ref{fig:network}. The particles connected with other particles via liquid bridge are joined with each other using a line. Each particle is equivalent to a vertex or node of a graph and the liquid bridge between two particles is equivalent to an edge of the graph. Using this imagery, Fig. \ref{fig:network} can be thought as an undirected graph $G = (V,E)$ where $V$ is a set of $n$ vertices (or nodes representing particles) and $E$ is a set of $m$ edges (representing particle-particle liquid bridge).} For an undirected graph, an edge is an unordered tuple $\{v_i,v_j\}$ where $v_i,v_j \in V$. A walk between node $i$ and $j$ represents a set of edges such that consecutive edges have a common node. Thus, a component of a graph $G = (V,E)$ is a subgraph $G_s = (V_s,E_s)$ such that $V_s \subseteq V$ and $E_s \subset E$ and there exists a walk between any two nodes $i,j \in V_s$. Thus, when the vertices represent particles and the edges represent presence of liquid-bridge, \textcolor{black}{then the simplistic definition of all particles interconnected by liquid bridges as a granule is identical to that of a component in a graph in network science terminology. Thus, Fig. \ref{fig:network} represents a graph with 22 nodes and 15 edges containing three components and 8 singlets.} 
 
 A weighted adjacency matrix of the graph defines the connectivity of the vertices as well as the strength of the connections, 
  \begin{equation}
	W_{ij} =  \begin{cases}
		w_{ij}, \; \text{if there is an edge between vertices}  \; i \; \text{and} \; j \\
		0, \; \text{otherwise}
	\end{cases}
\end{equation} 

 where $w_{ij}$ is the weight of the edge between nodes $i$ and $j$.
 For a network representation of a wet granular system in the present work, the vertices represent the particles while $w_{ij}$ corresponds to the liquid-bridge force $|\vec{F}_{lb,ij}|$ between particles $i$ and $j$. Thus
 
 \begin{equation}
 	W_{ij} =  \begin{cases}
 		|\vec{F}_{lb,ij}|, \ if |\vec{F}_{lb,ij}| > F_{lb,th} \\
 		0, \, otherwise
 	\end{cases}
 \end{equation}
 
\begin{figure*}[h]
 	\centering
 	\includegraphics[width=16cm]{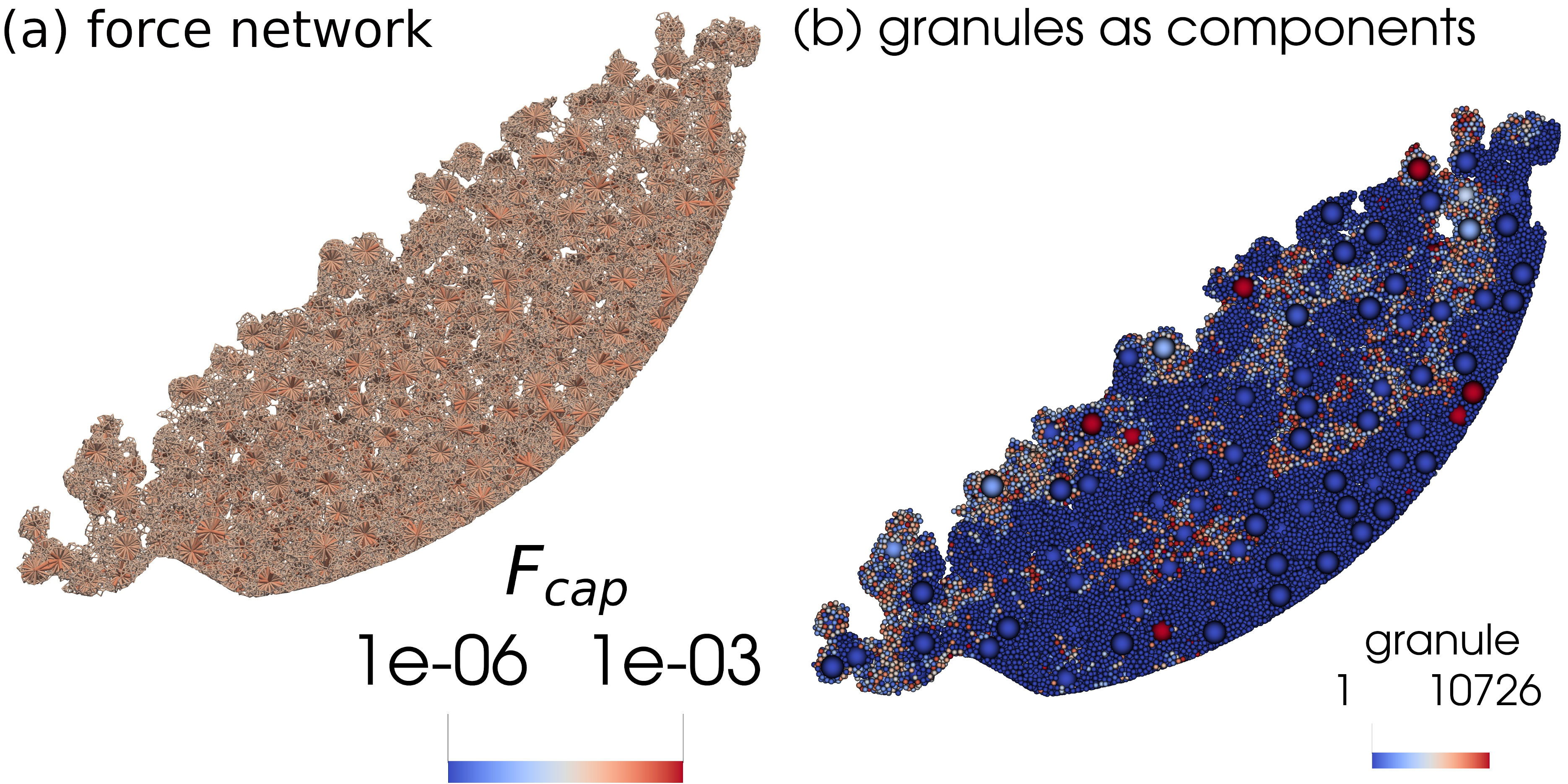}
 	\caption{Snapshots of (a) force network, (b) granules as components in the force network at threshold fraction $\zeta=0.6$. Here, fill fraction $f=0.16$ and liquid content $L=0.12$ during granulation.}
 	\label{fig:schematicComponent}
 \end{figure*}

 \textcolor{black}{where} $F_{lb,th}$ represents a threshold liquid-bridge force below which the liquid-bridge force between the particles is \textcolor{black}{neglected}. In the present work,  $F_{lb,th}$ is chosen such that the smallest $\zeta$ fraction of the liquid-bridge forces from the network are neglected. For example, if $\zeta=0.25$ then the smallest quartile of the liquid-bridge forces in the network were neglected. \textcolor{black}{Fig. \ref{fig:schematicComponent}(a) shows the liquid bridge force network and Fig. \ref{fig:schematicComponent}(b) shows the components measured for $\zeta=0.6$. Evidently, despite neglecting $60\%$ lower strength liquid bridges, a total of $10726$ granules are identified using this component definition.}
 
 \begin{figure*}[h]
 	\centering
 	\includegraphics[width=16cm]{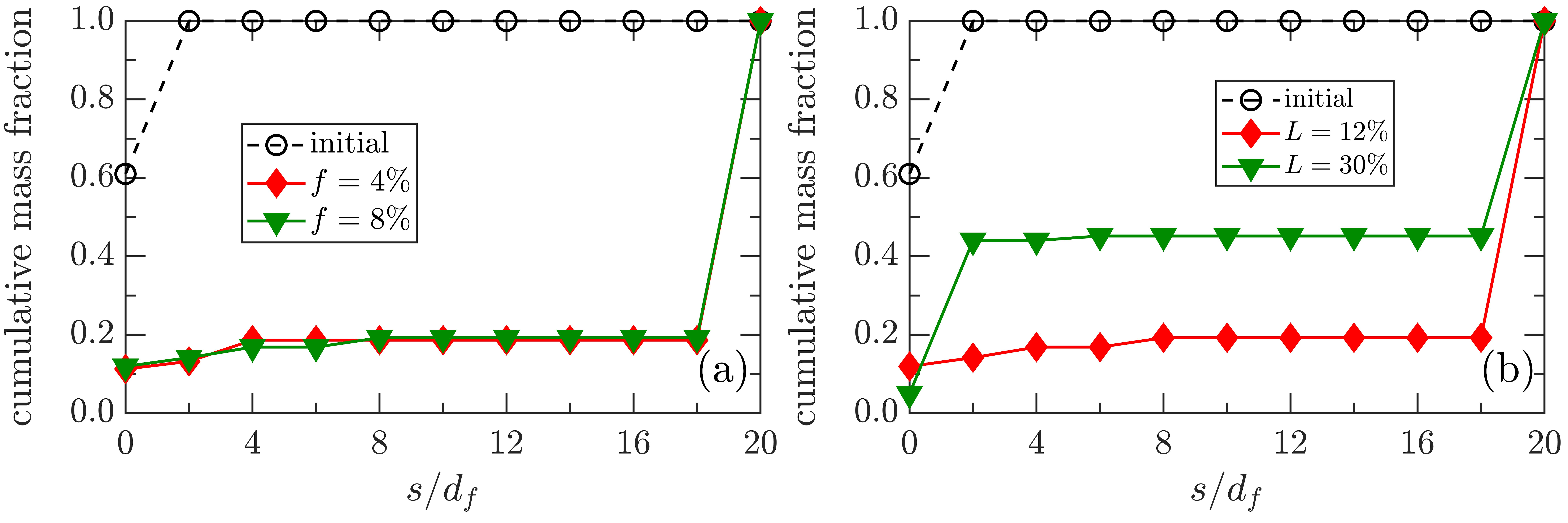}
 	\caption{Cumulative granule size distribution measured by component-detection method showing the effect of (a) fill level. Here, the liquid content $L=12\%$. and (b) liquid content. Here, the fill level $f=8\%$. Here, the threshold fraction $\zeta=0.6$.}
 	\label{fig:component}
 \end{figure*}
 
 Fig. \ref{fig:component}(a) shows the effect of fill level on the cumulative granule size distribution as measured using the component detection method \textcolor{black}{for two different fill levels of $4\%$ and $8\%$. The flat plateau region observed in the cumulative mass fraction over the large range of scaled granule size confirms that the component definition of granule size does not identify \textcolor{black}{significant number of} granules in the size range $4 \leq s/d_f < 18$. Note that size $s/d_f = 4 $ corresponds to the large size seeds. Thus around $20\%$ granules of size less than the $2$ mm size seeds are identified using this method. In both cases nearly $80\%$ of the mass is of granules more than $8$ mm ($s/d_f > 16$)}. In contrast, the sieving \textcolor{black}{(Fig.\ \ref{fig:sieving}(a)) and liquid-settling (Fig.\ \ref{fig:schematicSettling}(a))} results show nearly no mass fraction on the 8 mm sieve. This difference is due to \textcolor{black}{the fact that the} component detection method does not distinguish between number of connections between two groups of particles. Thus, even a single liquid-bridge force (in the top $(1-\zeta)$ fraction) between two groups of particles makes them a single component. In effect, this leads to the component-detection method identifying very large groups of particles as a single granule. Importantly, the rightward-shift due to increasing fill level on the cumulative granule size distributions is not reproduced by the component detection method. This suggests the inadequacy of the component detection method in measuring granule size distribution for discrete element modeling of granulation.
 
  \begin{figure*}[h]
 	\centering
 	\includegraphics[width=16cm]{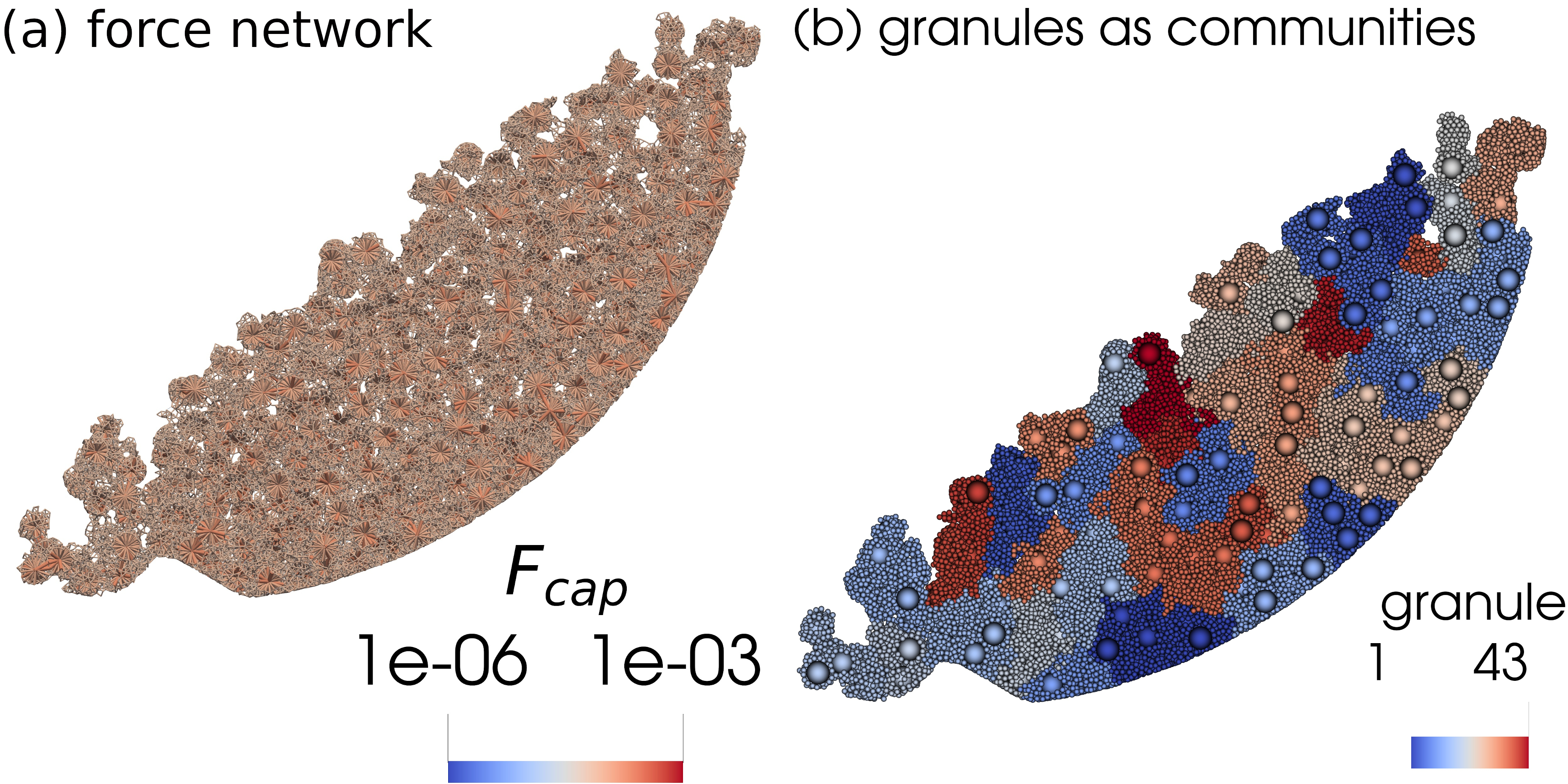}
 	\caption{Snapshots of (a) force network and (b) granules as communities in the force network at resolution parameter $\gamma=1.0$. Here, fill fraction $f=0.16$ and liquid content $L=0.12$ during granulation.}
 	\label{fig:schematicCommunity}
 \end{figure*}

  \begin{figure*}[h]
	\centering
	\includegraphics[width=16cm]{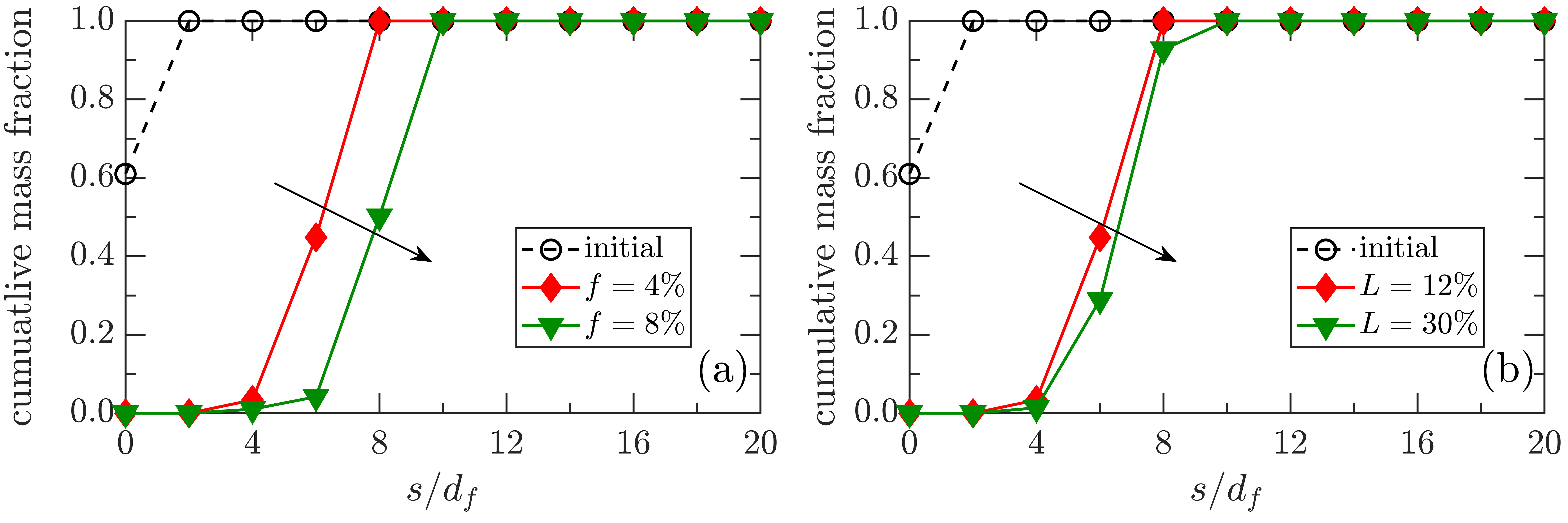}
	\caption{Cumulative granule size distribution measured by community detection showing the effect of (a) fill level. Here, the liquid content $L=12\%$ and (b) liquid content. Here the fill level $f=4\%$. Here the resolution parameter $\gamma=1.0$. }
	\label{fig:community}
\end{figure*}

Fig.\ref{fig:component}(b) shows the effect of liquid content on the cumulative granule size distributions as measured by the component detection method. \textcolor{black}{If this cumulative mass distribution with granule size is analysed in terms of mass collected on sieves of sizes $s$, with $s/d_f =0 $ corresponding to collection pan, the case with $12\%$ liquid content shows nearly $10\%$ by mass $0.5$ mm particles on the pan, and $\sim80\%$ of the mass is detected to be larger than the 8 mm size. In contrast, the case with 30\% liquid content shows only $\sim5\%$ of 0.5 mm particles on the pan. However, $\sim40\%$ mass of the granules is on the 1 mm sieve ($s/d_f =2$) and $\sim50\%$ mass is on the 8 mm sieve ($s/d_f \geq 16$).} This indicates that the case with higher liquid content ($30\%$) had a substantially lower proportion of large granules as compared to the case with lower liquid content ($12\%$). This result is in contradiction with well known effect of liquid content on the \textcolor{black}{granule size distribution}. Specifically, the \textcolor{black}{granule size distribution} shifts towards the right with increasing liquid content since it leads to production of larger granules. 

The results discussed above \textcolor{black}{confirm that the common method used in literature which considers all particles in contact via liquid bridge as a granule} is not appropriate for identifying granules and measuring \textcolor{black}{granule size distribution} via DEM for dense systems. \textcolor{black}{In fact even by using a minimum threshold for the liquid bridge force, the method fails to identify granule sizes appropriately. While the results are reported only for a particular threshold parameter $\zeta$, results at other threshold $\zeta$ values are also not satisfactory.} \textcolor{black}{Below, we propose a better method from the network-science literature, namely the community-det\-ection method, to identify granules and measure their size distributions}.  

  \subsubsection{Community-detection method}
 \label{subsubsec:4_3_2}
 
 A community is set of nodes which are more strongly and densely connected to each other than to any other set in the network \cite{newman_networks_2018}. A common method to detect communities is based on maximization of network property called modularity
 
  \begin{equation}
  	Q = \sum_{i,j}\left[W_{ij}-\gamma P_{ij}\right]\delta(c_i,c_j)
 \end{equation}
 
 where $Q$ is modularity, $\gamma$ is a resolution parameter, $P_{ij}$ is a matrix representing the expected edge weights connecting different pairs of nodes. \textcolor{black}{\begin{equation*}
     \delta(c_i,c_j) = \begin{cases}
         1, & \text{if } i \text{ and } j \text{ are in the same community}\\
         0, & \text{otherwise}
     \end{cases}
 \end{equation*} For very low values of $\gamma$, few large size communities are observed whereas for very high values of $\gamma$, large number of small size communities are observed. In this work, we use}
 
   \begin{equation}
 	P_{ij} = \frac{k_ik_j}{2m}
 \end{equation}
 
 where $k_i=\sum_{j}W_{ij}$ is the weighted degree of node $i$ and $m=\displaystyle\frac{1}{2}\sum_{ij}W_{ij}$ is the total weight of the network edges. \textcolor{black}{The modularity compares the number of edges inside community with that expected to happen by chance in a random network of the same size and \textcolor{black}{same number of connections for each node}. If the the nodes are much more connected to each other within a group than predicted by random chance, the modularity score is higher and community structure is strong. However, if the two do not differ significantly, no significant community structure exists and the system resembles to a  random network.}

 Thus, if we construct a network with particles as nodes with edges existing between particles with a minimum liquid-bridge force between them, then a granule can be defined as a component in the network. This is the common definition used in the granulation literature. Thus, in Fig.\ref{fig:network} the set of vertices $\left\{4,5,8,9,14,15,19,20 \right\}$ would be identified as a \textcolor{black}{single} granule by a component-detection method. On the other hand, a community-detection method would distinguish sets of vertices with stronger and denser connections and identify the sets $\left\{4,5,8,9\right\}$ and $\left\{14,15,19,20\right\}$ as separate granules.
 
 Fig. \ref{fig:schematicCommunity}(a) shows the network of capillary force among particles in a rotating drum and \textcolor{black}{Fig. \ref{fig:schematicCommunity}(b)} the communities detected using the network. Fig. \ref{fig:community}(a) shows the effect of fill level on the cumulative granule size distribution measured by the community detection method. We observe \textcolor{black}{that} increasing fill level leads to a rightward shift of the cumulative granule size distribution similar to the effect observed \textcolor{black}{via liquid-settling}. Fig. \ref{fig:community}(b) shows the effect of liquid content on the cumulative granule size distribution measured by the community-detection method. We \textcolor{black}{again} observe a rightward shift in the cumulative granule size distribution as expected from the results of sieving and liquid-settling.
 
The results above demonstrate that the component-detection method, \textcolor{black}{commonly employed in literature,} is ineffective in identifying appropriate groups of particles as granules for drum granulation. \textcolor{black}{It is} not able to capture the effects of process parameters (fill level and liquid content) on the granulation process. On the other hand, the community-detection method \textcolor{black}{is better} to identify granules which \textcolor{black}{compared} well with granules separated by liquid-settling. \textcolor{black}{It appears that} the cumulative granule size distributions measured by the community-detection method are able to capture the well-known effects of the material properties and process parameters on the granulation process.

  \begin{figure}[h]
	\centering
	\includegraphics[width=8cm]{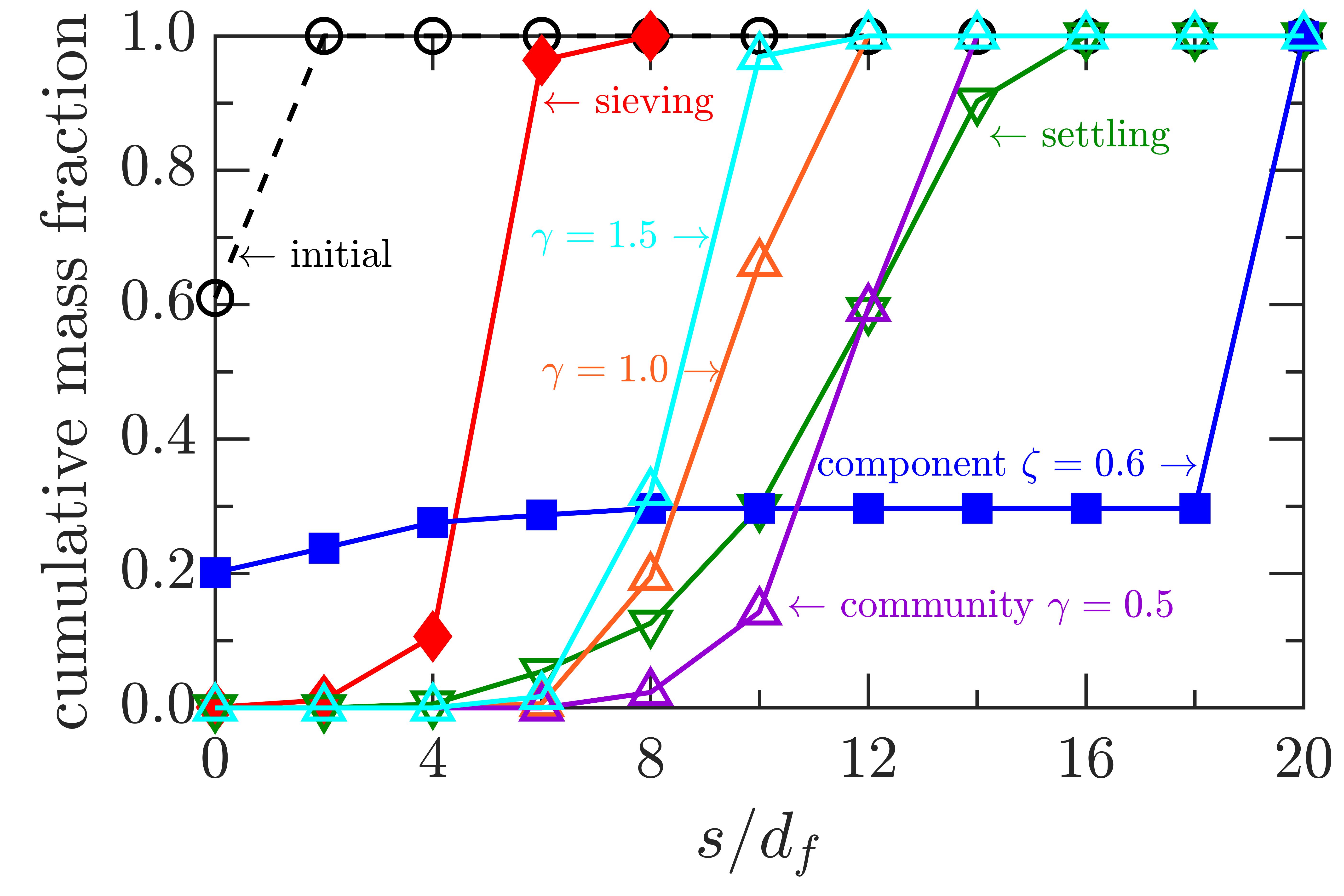}
	\caption{Cumulative granule size distribution measured by different methods. Here, the liquid content $L=12\%$ and the fill level $f=16\%$. The resolution parameter for community-detection method $\gamma \in \left\{0.5, 1.0, 1.5\right\}$. The threshold fraction for component-detection method $\zeta=0.6$.}
	\label{fig:comparsion}
\end{figure}

Fig. \ref{fig:comparsion} shows the cumulative granule size distribution measured by the component-detection method and community-detection method as compared against sieving and settling. We see that the granule size distribution measured by the component-detection method is very different from the other methods due to the limitations discussed above. \textcolor{black}{The} shape of the cumulative granule size distribution as measured by sieving, liquid-settling and community-detection are quite similar. We also see that by varying the resolution parameter $\gamma$ the cumulative granule size distribution measured by the community-detection method can be fitted with more conventional methods like sieving and liquid-settling. 

\section{Conclusion}
We investigate the effectiveness of the commonly-used component-detection in identifying granules for discrete element modeling of granulation. \textcolor{black}{The distribution of granule mass as predicted by the component-detection method shows that the method identifies granules which are \textcolor{black}{either} too large (nearly encompassing the entire drum mass) or too small (one to a few primary particles).} We propose an improved method based on the idea of community-detection from the network science literature. \textcolor{black}{The granules identified by the community-detection method were very similar in their size} to granules separated using conventional methods like liquid-settling. The effectiveness of the proposed method for the study of granulation \textcolor{black}{is} demonstrated by computing the cumulative granule size distribution. Using the proposed method, we \textcolor{black}{are} able to reproduce \textcolor{black}{the well-known effects of} fill level and liquid content on the granulation performance.

\section*{Declaration of competing interests}
The authors declare that they have no known competing financial interests or personal relationships that could have appeared to influence the work reported in this paper.

\section*{Acknowledgement}
AT, JC and JK acknowledge the infrastructure support received from DST-SERB (CRG/2022/005725) for performing this work. SJ acknowledges the Junior Research Fellowship from DST-SERB (CRG/2022/005725). In addition, the support and the resources provided by PARAM Sanganak under the National Supercomputing Mission, Government of India at the Indian Institute of Technology, Kanpur
are gratefully acknowledged.	

\section*{Data availability}
Data will be made available on request.	
	
		\appendix
		
		\bibliographystyle{elsarticle-num}
		\bibliography{bibliography.bib}
	
\end{document}